\documentclass[prb,aps,twocolumn,showpacs]{revtex4}
\usepackage{graphicx}
\usepackage{wasysym}
\usepackage{dcolumn} 
\usepackage{bm, bbm}      
\usepackage{curves}
\usepackage{epic}
\usepackage{epsf, times}
\usepackage{subfigure}
\usepackage{eufrak}

\newcommand{\bone}{\mathbbm{1}}

\newcommand{\rhobar}{\bar{\rho}}
\newcommand{\Sbar}{\bar{S}}

\newcommand{\Pbar}{\bar{P}}

\newcommand{\sigmab}{\mbox{\boldmath $\sigma $}}

\newcommand{\beq}{\begin{equation}}
\newcommand{\eeq}{\end{equation}}
\newcommand{\bea}{\begin{eqnarray}}
\newcommand{\eea}{\end{eqnarray}}
\newcommand{\nn}{\nonumber}
\newcommand{\bei}{\begin{itemize}}
\newcommand{\eei}{\end{itemize}}
\newcommand{\etal}{{\em et al.}}

\newcommand{\bq}{{\bf q}}

\newcommand{\bs}{{\bf S}}
\newcommand{\br}{{\bf r}}
\newcommand{\bn}{{\bf n}}

\newcommand{\tS}{\theta_S}
\newcommand{\tI}{\theta_I}
\newcommand{\pS}{\phi_S}
\newcommand{\pI}{\phi_I}
\def\sta#1{|#1\rangle}
\def\cs{{\cal S}}
\def\ct{{\cal T}}
\def\cm{{\cal M}}
\def\cn{{\cal N}}
\def\ce{{\cal E}}
\def\cg{{\cal G}}

\def\tit#1#2#3#4#5{{#1}{\bf #2}, #3 (#4)}

\def\prl{Phys.\ Rev.\ Lett.\ }

\def\prb{Phys.\ Rev.\ B\ }

\begin{document}

\title{Entanglement Skyrmions in multicomponent quantum Hall systems}

\author{B. Dou\c{c}ot,$^1$ M. O. Goerbig,$^2$ P. Lederer,$^2$ and
  R. Moessner$^3$} 
\address{$^1$Laboratoire de Physique Th\'eorique et Hautes Energies, CNRS UMR
  7589, Universit\'e Paris 6 et 7, Paris\\
$^2$Laboratoire de Physique des Solides, CNRS UMR 8502, Universit\'e Paris
  Sud, Orsay\\
$^3$Max-Planck-Institut f\"ur Physik komplexer Systeme, 01187 Dresden} 

\date{\today}

\begin{abstract}
We discuss charged topological spin textures in quantum Hall ferromagnets in
which the electrons carry an pseudospin as well as the usual spin degree of
freedom, as is the case in bilayer GaAs or monolayer graphene samples. We
develop a theory which treats spin and pseudospin on a manifestly equal footing,
which may also be of help in visualizing the relevant spin textures. We in
particular consider the entanglement of spin and pseudospin in the presence of
realistic anisotropies. An entanglement operator is introduced which generates
families of degenerate Skyrmions with differing entanglement properties. We
propose a local characterization of the latter, and touch on the role entangled
Skyrmions play in the nuclear relaxation time of quantum Hall ferromagnets.  
\end{abstract}

\pacs{73.43.Lp, 73.21.-b, 81.05.Uw}

\maketitle

\section{Introduction}
Quantum Hall systems with supplementary degrees of freedom are fascinating
objects of study, exhibiting phenomena such as the superfluid behaviour in
quantum Hall bilayers at total filling factor 
$\nu=1$.\cite{eisenstein,wen,SF2,SF3,SF4,SF5,SF6,SF7} In
those systems, the superfluid  ground state  is isomorphous to a quantum Hall
monolayer ferromagnetic ground state  at the same filling factor, but in terms
of the $S=1/2$ pseudospin degree of freedom encoding which layer an electron finds
itself in. A similar pseudospin ferromagnet is surmised to occur in a graphene
monolayer, where the pseudospin degree of freedom is analogously connected to the
two degenerate mid band valleys.\cite{nomura,GMD,alicea}

In quantum Hall ferromagnets at $\nu=1$, the lowest charged excited states are
topological defects, called Skyrmions,\cite{skyrme} which in turn are
fascinating in their own right. Indeed they are spin  textures carrying a
topological quantum number and which necessarily carry a quantized net
charge. The properties of Skyrmions in quantum Hall (QH) physics have been
studied extensively since their theoretical prediction\cite{sondhiskyr} and
experimental confirmation\cite{barretNMR} in the mid 1990s. These charged
topological spin textures are due to the strong electronic correlations once
the low-energy excitations are restricted to a single Landau level (LL). They
can exist as quasiparticles of a ferromagnetic ground state near a LL filling
factor $\nu=1$, where $\nu=n_{el}/n_B$ is the ratio between electronic density
$n_{el}$ and $n_B={eB}/{h}=1/2\pi l^2_B$ that of the flux quanta
$h/e$ threading an area of the two-dimensional (2D) electron gas at a
particular magnetic field $B$ (here, $l_B=\sqrt{\hbar/eB}$ is the magnetic
length). 

In bilayer QH systems, in contrast to the intrinsic electronic spin $1/2$ in
monolayer QH systems, the effective interaction becomes pseudospin-dependent
due to the different interaction strength between electrons in the same layer
and that between particles in different layers.\cite{moon,GirvinHouches} It is
indeed energetically unfavorable to have all electrons in the same layer
because of a finite amount of charging energy per particle. This results in an
easy-plane ferromagnetic ground state, i.e. in which the pseudospin
magnetization is constrained in the $xy$-plane. Bilayer ferromagnets have,
thus, an easy-plane [U(1)] symmetry.

Topological excitations of easy-plane ferromagnets in bilayer QH systems are
conventionally described in terms of merons classified by their charge ($\pm
1/2$) and their vorticity ($\pm 1$).\cite{moon} As each meron has a
logarithmically divergent energy, one needs two merons with opposite vorticity
to make a topologically stable  excitation of finite energy (bimeron), which
may also be viewed as a pseudospin Skyrmion.  

Besides the physical spin in monolayer and the layer pseudospin in bilayer QH
systems, valley indices can be encoded by pseudospin degrees of freedom. The
indirect-gap semiconductors Si and AlAs are well-studied 
instance,\cite{SiSC,AlAs} as is
the pair of Dirac points in graphene.\cite{Antonio}

One particularly interesting situation arises when the internal Hilbert
space comprises both the spin and the additional pseudospin. This combination
can in principle lead to an internal space with a high symmetry, such as SU(4)
for a combination of a spin and pseudospin pair of SU(2)'s. The properties of
SU(4) Skyrmions were considered by Ezawa\cite{ezawa1} and that of general
SU($N$) Skyrmions for $N>2$ by Arovas {\sl et al.}\cite{arovas} SU(4) Skyrmions
have recently been revisited by Yang {\sl et al.} in the context of
graphene.\cite{yang} 

However, the Hamiltonian generically contains anisotropy terms violating the
full symmetry. In the simplest situation of a Zeeman field, the symmetry
$\sigma^z\leftrightarrow-\sigma^z$ is removed. Such symmetry-breaking terms can
in principle also be manipulated, for example the effective $g$-factor
determining the strength of the Zeeman coupling can be tuned by applying
external pressure thanks to spin-orbit coupling. The energetics of the
resulting topological excitations has been an active field of study in
recent years.\cite{Cote91,Cote97,Ezawa02,ezawa2,tsitsishvili,Bourassa06,Cote07} 

In this paper, we consider the properties of SU(2)$\times$SU(2) Skyrmions in
the presence of realistic anisotropy terms. We do this from a perspective of
entanglement between the spin and pseudospin degrees of freedom. As our point of departure, 
we parametrise the general SU(4) Skyrmion using a Schmidt
decomposition.\cite{schmidt} Our parametrisation is  manifestly symmetric 
between the two SU(2) copies arising from spin and pseudospin. In addition, a
third (`entanglement') spinor appears. 
These features turn out to be of help in 
visualizing the nature of different types of SU(4) Skyrmions, for instance by
providing a simple picture for the relative entanglement between spin and pseudospin.

With this in hand, we compute the Berry
connection and then display in a transparent way the properties of entanglement
textures. (The topological stability of the SU(4) Skyrmion becomes, strictly
speaking, apparent only in the associated CP$^3$ description.) Crucially, we
show that there exist entire U(1) families of degenerate Skyrmions which differ
by their degree of entanglement. 

This in particular leads to the distinction of two different topological
excitations, both of which have zero expectation value of transverse spin but
in only one of which the spin degree of freedom is involved. At some special
``maximally entangled'' points, the `order parameters' $\langle \vec{\sigma}
\rangle$ and $\langle \vec{\tau} \rangle$ may even vanish altogether. 

This treatment offers some further insights into the magnetic properties of
multi-component systems, which have been studied experimentally in recent
years, in the framework of nuclear magnetic resonance (NMR) experiments on
bilayer QH systems.\cite{Spielman05,Kumada05} In particular, we briefly discuss
how entangled excitations affect the spin degrees of freedom measured in NMR
experiments.  
 
The paper is organized as follows. In Sec. II, we introduce the basic model and
the parameterization of the spin-pseudospin degrees of freedom. Spin-pseudospin
entanglement is studied in Sec. III, where we propose a measure of such
entanglement, the properties of which are described in Sec. IV. Sec. V reviews
the relevant energy scales of bilayer and graphene QH systems, where
entanglement Skyrmions may be physically relevant. The dynamical properties of
entangled spin-pseudospin textures and their consequences for NMR measurements are
discussed in Sec. VI. 

\section{Model and parameterization}
We study a two-dimensional electron system (2DES) in the QH regime, where the
kinetic energy is quenched. We consider two internal degrees of freedom which
provide the multicomponent nature of the 2DES. Firstly, the spin degree of
freedom, represented by the Pauli matrices $\sigma^a$, where
$a=x,y,z$. Secondly, the pseudospin, represented by a second set of Pauli
matrices, $\tau^\mu$, with $\mu=x,y,z$. For concreteness, it may help to have
the bilayer index in mind when considering the pseudospin.

\subsection{Parameterization}

One may typically describe a texture at overall filling factor $\nu=1$ by a
Hartree-Fock state of the form
\bea
|\Phi\rangle & = &
\prod_{X}\left[w_1(X)c^{+}_{X\uparrow,t}+w_2(X)c^{+}_{X\uparrow,b} +
  w_3(X)c^{+}_{X\downarrow,t} \right. \nn \\ 
& & \mbox{} \left.+ w_4(X)c^{+}_{X\downarrow,b}\right]|0\rangle
\eea
where $X$ labels orbital states in the plane restricted to the lowest LL.
Internal states are labeled according to the projection of the spin along the
external magnetic field ($\uparrow$ or $\downarrow$) and the layer ($t$ or
$b$). We assume that the orbital states are each fairly well localized (on a
typical scale given by the magnetic length) and mutually orthogonal. The
explicit construction of such a basis for any Landau level is given for
instance in Ref. \onlinecite{Rashba97} and has been used to describe  the
bilayer system by a lattice SU(4) model.\cite{Burkov02} With such a choice,
$X\equiv\br$ can be viewed as the the location of the center of a given Wannier
orbital on a von Neumann lattice. We will always assume that the texture varies
slowly on the scale of the magnetic length, so that the $w_{i}$ ($1 \leq i \leq
4$) may be regarded as smooth complex functions of a continuous position
variable. We will also assume these local four-component spinors to be
normalized, that is $\sum_{i=1}^{4}|w_i(\br)|^{2}=1$. To get a more direct
physical interpretation of such texture, it is rather natural to express the
internal state at site $\br$ in a way that treats both spin and pseudospin
degrees of freedom on an equal footing. This is achieved by the Schmidt
decomposition\cite{schmidt}  
\bea
\nn
|\Psi(\br)\rangle &=& \cos\frac{\alpha}{2}|\psi_S\rangle|\psi_I\rangle
\\
&&+\sin\frac{\alpha}{2} e^{i\beta}|\chi_S\rangle|\chi_I\rangle\ 
\label{Schmidt}
\eea
where $\alpha$ and $\beta$ are functions of $\br$, and the local two-component
spinors $|\psi_{S}\rangle$, $|\chi_{S}\rangle$, $|\psi_{I}\rangle$, and
$|\chi_{I}\rangle$ are constructed according to 
\bea
|\psi\rangle&=&
\left( 
{\cos\frac{\theta}{2}}\atop{\sin\frac{\theta}{2}e^{i\phi}}
\right)\\
|\chi\rangle&=&\left(
{-\sin\frac{\theta}{2}e^{-i\phi}}\atop{\cos\frac{\theta}{2}}
\right)\ .
\eea
Here, $\theta$ and $\phi$ are the usual polar angles defining a vector
\bea
\bn(\theta,\phi)=\left(\sin\theta\ \cos\phi,\sin\theta\
\sin\phi,\cos\theta\right)\ . 
\eea
We will denote the pair for spin (pseudospin) by $\tS,\pS$ ($\tI,\pI$),
respectively. Note that the above parameterization for the two component
spinors is not unique, because for a given classical unit vector
$\bn(\theta,\phi)$ we can multiply both $|\psi\rangle$ and $|\chi\rangle$ by
global phases that may depend on $\br$. We have chosen the parameterization in
order to minimize the occurrence of phase singularities -- they appear only
when one of the angles $\theta_S$, $\theta_I$, or $\beta$ is equal to $\pi$.

With these notations, the general four-component local spinor
(Eq.~\ref{Schmidt}) has the form
\begin{widetext}
\bea
|\Psi(\br)\rangle=\left(\begin{array}{c}
w_1 \\ w_2 \\ w_3 \\ w_4
\end{array}
\right) = 
\left(\begin{array}{ccc}
\cos\frac{\theta_S}{2}\cos\frac{\theta_I}{2}\cos\frac{\alpha}{2}
&+& \sin\frac{\theta_S}{2}\sin\frac{\theta_I}{2}\sin\frac{\alpha}{2}
e^{i(\beta-\phi_S-\phi_I)}\\
\cos\frac{\theta_S}{2}\sin\frac{\theta_I}{2}\cos\frac{\alpha}{2}e^{i\phi_I}
&-&\sin\frac{\theta_S}{2}\cos\frac{\theta_I}{2}\sin\frac{\alpha}{2}
e^{i(\beta-\phi_S)}\\
\sin\frac{\theta_S}{2}\cos\frac{\theta_I}{2}\cos\frac{\alpha}{2}e^{i\phi_S}
&-& \cos\frac{\theta_S}{2}\sin\frac{\theta_I}{2}\sin\frac{\alpha}{2}
e^{i(\beta-\phi_I)}\\
\sin\frac{\theta_S}{2}\sin\frac{\theta_I}{2}\cos\frac{\alpha}{2}e^{i(\phi_I+\phi_S)} 
&+& \cos\frac{\theta_S}{2}\cos\frac{\theta_I}{2}\sin\frac{\alpha}{2}
e^{i\beta}
\end{array}\right)\ .
\label{spinor}
\eea
\end{widetext}

We thus have six parameters ($\tS,\pS,\tI,\pI,\alpha,\beta$), as expected for
four complex components minus a global phase and the overall normalization. All
of these can vary as a function of the spatial coordinate $\br=(x,y)$ in the
$xy$-plane. 

\begin{figure}
\centering
\includegraphics[width=8.5cm,angle=0]{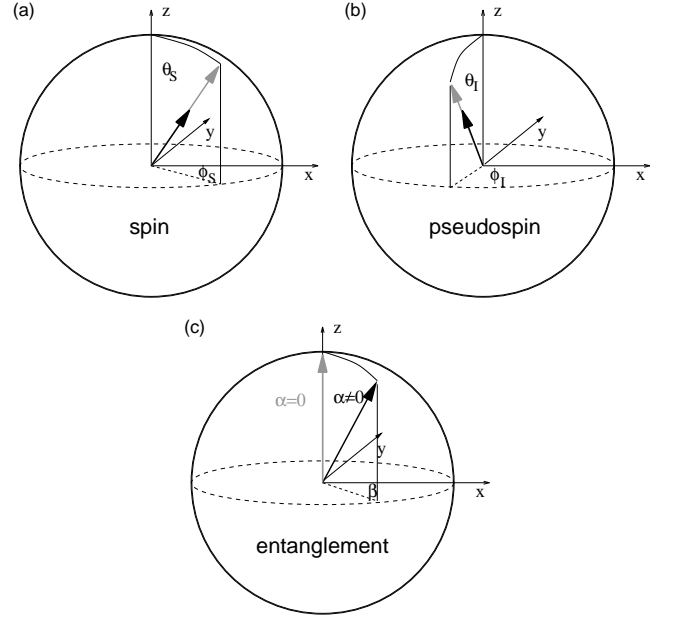}
\caption{\footnotesize{Bloch spheres for entangled spin-pseudospin systems. 
Bloch sphere for the spin {\sl (a)}, pseudospin {\sl (b)}, and a third type
of spin representing the entanglement {\sl (c)}. In the case of spin-pseudospin
entanglement ($|\cos\alpha|\neq 1$), the (pseudo)spin-magnetizations explore
the interior of their spheres, respectively (black arrows). 
}} 
\label{fig01}
\end{figure}

The great advantage of factorizing the wavefunction in such a manner is that
we can read off directly the reduced density matrices for the spin
and pseudospin sectors:

\bea
\nn
\rho_S &=& {\rm Tr}_I\left(|\psi\rangle\langle \psi|\right)=
\cos^2\frac{\alpha}{2}|\psi_S\rangle\langle \psi_S|+\sin^2\frac{\alpha}{2}
|\chi_S\rangle\langle \chi_S|\ ,\\
\nn
\rho_I &=& {\rm Tr}_S\left(|\psi\rangle\langle \psi|\right)=
\cos^2\frac{\alpha}{2}|\psi_I\rangle\langle \psi_I|+\sin^2\frac{\alpha}{2}
|\chi_I\rangle\langle \chi_I|\ ,
\eea
for the spin and the pseudospin, respectively. This yields 
\beq\label{LSdens}
m_{S}^{a}={\rm Tr}(\rho_S S^{a})=\cos\alpha\langle\psi_S|S^{a}|\psi_S
\rangle=\cos\alpha \, n^a(\theta_S,\phi_S)
\eeq
for the local spin and
\beq\label{LPdens}
m_{I}^{\mu}={\rm Tr}(\rho_I P^{\mu})=\cos\alpha\langle\psi_I|P^{\mu}|\psi_I
\rangle=\cos\alpha \, n^{\mu}(\theta_I,\phi_I)
\eeq
for the local pseudospin density. Notice that for the case $\alpha\neq 0$ or $\pi$
(i.e. $\cos^2\alpha<1$), the local (pseudo)spin densities are no longer normalized,
but are of length $|{\bf m}_{S/I}|^2=\cos^2\alpha$. Thus, in a semiclassical
picture, the (pseudo)spin dynamics is no longer restricted to the surface of
the Bloch  sphere, but explores the entire volume enclosed by the sphere
(Fig. \ref{fig01}).

\subsection{Topological densities}

For an SU(2) symmetry, spin textures in the 2D plane may be classified
according to which homotopy class they belong. First, one needs to map the $xy$
plane, by stereographic projection, onto the $S^2$ sphere. For an XYZ
ferromagnet, the most general spin texture parameterization maps on the surface
of the Bloch sphere $S^2$. The mapings $S^2 \rightarrow S^2$, which can be
continuously deformed into one another, may be collected into homotopy
classes.\cite{mermin} The latter form a homotopy group $\pi_2(S^2)= \mathbbm{Z}$. The
elements of $\mathbbm{Z}$, which are integers $Q$, characterize different
topological sectors. Spin waves, for instance, belong to the topological
sector with topological charge $Q=0$. Skyrmions carry a non-zero topological
charge $Q\neq 0$. In the QH ferromagnet, a spin texture generates a charge
density connected to the former through the so-called topological density. The
topological density arises because an electron travelling in a spin texture
acquires a Berry phase analogous  to an Aharonov-Bohm phase. The total charge,
expressed in units of the electron charge,  of a spin texture in a QH state
with Laughlin filling factor $\nu=1/(2m+1)$ is given by $\nu Q$.\cite{moon,GirvinHouches}

Next, we transpose these ideas to the larger SU(4) spin-pseudospin symmetry. 
With the help of the Schmidt decomposition, one may calculate the charge
density induced by a general spin-pseudospin texture. When the internal states
are described by a single SU(2) spin, the local density is simply given by the
topological density of the associated director field, that is
\bea
\rho_{top}=\frac{\epsilon^{ij}}{8\pi}\bn(\theta,\phi)\cdot
[\partial_i\bn(\theta,\phi)\times\partial_j\bn(\theta,\phi)], 
\eea
where $\epsilon^{ij}$ is the antisymmetric tensor, $\epsilon^{xy}=-\epsilon^{yx}=1$ and 
$\epsilon^{ii}=0$.
To generalize this to the present case, it is sufficient to note that the
spatial variation of $|\Psi(\br)\rangle$ due to the texture induces a Berry
phase, which is encoded by the Berry connection,
\beq
{\cal A}(\br)=\frac{1}{i}\langle\Psi|\nabla \Psi\rangle .
\eeq 
The local charge density carried by the texture is then the 2D curl
${\cal B}(\br)=[\partial_x{\cal A}_y(\br)-\partial_y{\cal A}_x(\br)]/(2\pi)$ 
of the fictitious vector potential ${\cal A}(\br)$. After some simple algebra, we
get 
\bea
{\cal A}(\br) &=&\sin^2\frac{\alpha}{2}\nabla\beta+ \nn \\
& &\cos\alpha\left(\sin^2\frac{\tS}{2}\nabla\pS+
\sin^2\frac{\tI}{2}\nabla\pI\right),
\eea
the curl of which  is, thus,
\bea
{\cal B}(\br)&=&
\cos\alpha\left\{\rho_{top}[\bn(\tS,\pS)]+\rho_{top}[\bn(\tI,\pI)]\right\}+
\nonumber\\
&&\rho_{top}[\bn(\alpha,\beta)]+\\
&&\sin^2\frac{\tS}{2}\rho_{top}[\bn(\alpha,\pS)]+
\sin^2\frac{\tI}{2}\rho_{top}[\bn(\alpha,\pI)]\ .\nonumber
\eea

\section{Spin-pseudospin entanglement}
\label{entanglement}
\subsection{A measure of entanglement}
The above parameterization makes it clear that there are Skyrmions in which
spin and pseudospin can be perfectly entangled. This is most simply the case
where for instance  
$|\psi_{S}\rangle=\left({1}\atop{0}\right)$,
$|\chi_{S}\rangle=\left({0}\atop{1}\right)$,
$|\psi_{I}\rangle=(1/\sqrt{2})\left({1}\atop{1}\right)$ and
$|\chi_{I}\rangle=(1/\sqrt{2})\left({-1}\atop{1}\right)$ for all
$\br$, and the Skyrmion is a texture in $\alpha(\br)$ and $\beta(\br)$
only [which can take the same form as $\tS,\pS$ in a pure spin Skyrmion. This
  form is explicited in Eq. (\ref{skyrmion_shape}) below].  In this case, it is
easy to see, e.g., that $\sigma^x\equiv0$ throughout as this operator is
off-diagonal in this basis but has no matrix elements between
$|\psi_S\rangle|\psi_I\rangle$ and $|\chi_S\rangle|\chi_I\rangle$. At the same
time, $\sigma^z$ has the same profile as in a standard spin Skyrmion, {because
  its expectation value is simply given by $\cos\alpha(\br)$}.

The vanishing of the expectation value of the transverse components of the spin
is thus a consequence of their being entangled with the pseudospin degree of
freedom. To quantify this entanglement, we propose to use the following
measure: 
\bea\label{eq:Xi}
\Xi:=1-\sum_a\langle\sigma^a\rangle^2=1-\sum_{\mu}\langle\tau^{\mu}\rangle^2\ .
\eea
Eq. (\ref{eq:Xi}) is easily expressed from the Schmidt decomposition Eq. (\ref{Schmidt})
as
\bea
\Xi=\sin^{2}\alpha
\eea

As a result, for the abovementioned entanglement texture, which may be
visualized on a third Bloch sphere [Fig. \ref{fig01}(c)], local entanglement is
minimal ($\Xi=0$) when $\sin\alpha=0$, that is $\alpha=0$ or $\alpha=\pi$,
corresponding to the center of the texture or to the points at
infinity. Entanglement is maximal ($\Xi=1$) when $\alpha=\pi/2$, that is at the
a distance from the origin given by the overal size $\lambda$ of the texture. 
Note that at a maximally entangled point, the moduli of both spin and
pseudospin vanish, as may be seen from Eqs. (\ref{LSdens}) and (\ref{LPdens}).

Alternatively, for the spinor defined in Eq. (\ref{spinor}), 
\bea
\Xi=4|w_1w_4-w_2w_3|^2 \ .
\eea 
For a factorizable state, in which the wavefunction can assume the form of an
outer product between spin and pseudospin parts,
$|\Psi_{\lambda\mu}\rangle=u_\lambda v_\mu$, one immediately finds
$\Xi=0$. Note that conversely, a vanishing $\Xi$ implies that we have a
factorizable state.  For a completely entangled state,
e.g. $w_1=w_4=1/\sqrt{2}, w_2=w_3=0$, we have $\Xi=1$.

\subsection{Degenerate texture families in the presence of spin anisotropy}

In the presence of a Zeeman field, the SU(2) symmetry of the spin degree of
freedom gets reduced to a U(1) symmetry of rotations around the $z$-axis (along
which we take the external field to be aligned). The generator of this symmetry
is $\sigma^z$. We furthermore consider, for illustration reasons, that the
SU(2) pseudospin symmetry is likewise broken to U(1), and we choose the
$z$-axis such that the rotations generated by $\tau^z$, the $z$-component of
the pseudospin, commute with the underlying Hamiltonian. We discuss, in Sec. V,
some physical examples, in the context of bilayer quantum Hall systems and
graphene, where this situation arises.

Given that the underlying model Hamiltonian commutes with $\tau^z$ and with
$\sigma^z$, it also commutes with their product, $\sigma^z\tau^z$. This means
that a textured state $\sta\ct$ belongs to a degenerate family of textures
$\sta{\ct(\gamma)}=\exp(i \gamma\tau^z\sigma^z)\sta{\ct}$. {The most
  interesting property of this latter symmetry operation is its ability to
  transform a factorizable state into an entangled one. Indeed, it acts on the
  four component spinors according to} $w_{1,4}\rightarrow w_{1,4}\exp(i
\gamma)$ and $w_{2,3}\rightarrow w_{2,3}\exp(-i \gamma)$.  In the process,
$|w_1w_4-w_2w_3|^2
\rightarrow|w_1w_4|^2+|w_2w_3|^2-[w_1w_4\bar{w}_2\bar{w}_3\exp(4i\gamma)+c.c.]$, 
where the bar indicates complex conjugation. Therefore, as $\gamma$ varies,
$\Xi$ varies between $\Xi_{{min\atop{max}}}$ with
\bea
\Xi_{{min\atop{max}}}=4(|w_1w_4|\mp|w_2w_3|)^2\ . 
\eea

It is interesting to specify how one may, thus, create a maximally entangled
state. This corresponds to $\Xi_{max}=1$, or equivalently to $|w_1|=|w_4|$ and
$|w_2|=|w_3|$, because one may rewrite
\bea
\Xi_{max}&=&1-[(|w_1|+|w_4|)^2+(|w_2|+|w_3|)^2]\times\nonumber\\ 
&&\
[(|w_1|-|w_4|)^2+(|w_2|-|w_3|)^2]\ .
\eea 
Note that creation of a maximally entangled state starting from a factorized
one is possible when the four spinor components have the same modulus.  Another
limiting case is obtained when $\Xi_{min}=\Xi_{max}$, that is the action of the
$\sigma^{z}\tau^{z}$ generator does not change the degree of entanglement,
although the quantum state does change as the angle $\gamma$ varies. This
requires that at least one of the four spinor components vanishes. If the
initial state is factorizable, this translates into $w_1w_4=w_2w_3=0$.  

\section{Consequences of entanglement}

\subsection{Descendants of Skyrmions and bi-merons}

Let us now consider the effect of these {entangling transformations}
on the textures {that are most relevant for the physics of the QH
effect in a bilayer system at a total filling factor $\nu$ close
to 1.} We choose, as a starting point, 
a `bimeron' texture, $\cm$, in which the
spin is constant along the magnetic field and the pseudospin has a
topologically non-trivial texture [see Eq.~\ref{eq:cm} below]. 
Due to the easy-plane anisotropy, the pseudospin is oriented  
in an arbitrary direction within the $xy$-plane far from the meron
core, whereas in the core region it is oriented in the $\pm z$-direction.
There are furthermore two possible vorticities, such that one may
distinguish between four meron types, each of which has half a 
charge.\cite{moon} A meron in itself costs a logarithmically large 
amount of energy due to the vorticity and must therefore be accompanied
by a second meron of opposite vorticity, in order to obtain a 
pseudospin texture of finite energy. If both merons have the same
topological charge, one obtains a bimeron with an overall topological
charge of $\pm 1$, otherwise the bimeron is topologically connected to the 
(ferromagnetic) ground state.
The action of $\tau^z$ on this state is simply to
rotate the texture in the easy plane of the pseudospin, which is a
global symmetry operation. The action of $\sigma^z$ is trivial, as all
spins point in the $z$-direction and hence a rotation about this axis
has no effect. Therefore, the combined effect of $\tau^z\sigma^z$ is
the same global rotation as that effected by $\tau^z$ alone,
and no entanglement is generated.  This is in agreement with
the general discussion given above, because we have for instance
$w_3=w_4=0$ if all the spins are up, which clearly satisfies
$w_1w_4=w_2w_3=0$.

Next, we consider Sondhi {\sl et al.}'s original spin-Skyrmion texture
$\sta\cs$, in which the pseudospin lies in a fixed direction in the
$xy$-plane (the $x$-direction, say). The effect of a rotation generated by
$\tau^z$ is simply to rotate the pseudospin uniformly about the $z$-axis
in pseudospin space.  The effect of $\sigma^z$ rotations is a global
rotation of the spin texture about the $z$-axis in spin space.  Now, the
crucial observation is that the effect of the combination
$\tau^z\sigma^z$ is no longer just a trivial rotation.  Rather, the
operation generates a texture in which spin and pseudospin are
entangled.  To be more specific, we choose the pure spin texture
as
\beq
\sta\cs=\left( 
{\cos\frac{\theta(\br)}{2}}\atop{\sin\frac{\theta(\br)}{2}e^{i\phi(\br)}}
\right)_{S}\otimes\frac{1}{\sqrt{2}}\left({1}\atop{1}\right)_{I}
\label{spinText}
\eeq
where $\br=(x,y)$ denotes the location in the plane to which the local spinor
$\sta\cs$ is associated. 
The second term in Eq. (\ref{spinText}) precisely 
means that the pseudospin is  homogeneously oriented in the
$x$-direction, whereas the first one describes the (pure) spin texture.
For a usual Skyrmion located at the origin,
we have
\beq
\tan\frac{\theta(\br)}{2}e^{i\phi(\br)}=\frac{x+iy}{\lambda}
\eeq
where $\lambda$ has the dimension of a length and corresponds to the
spatial extension of the Skyrmion.
In this fully separable state, the expectation value of the spin operator reads: 
\bea
\label{skyrmion_shape}
\langle\sigma^{x}\rangle & = & \frac{2\lambda x}{\lambda^{2}+x^{2}+y^{2}}\, ,
\\ 
\langle\sigma^{y}\rangle & = & \frac{2\lambda y}{\lambda^{2}+x^{2}+y^{2}}\, ,
\\ 
\langle\sigma^{z}\rangle & = &
\frac{\lambda^{2}-x^{2}-y^{2}}{\lambda^{2}+x^{2}+y^{2}}\, , 
\eea
which shows that the complex number $(x+iy)/\lambda$ is simply given by the
stereographic projection from the south pole of the unit vector
$\langle\sigmab\rangle$ onto the complex plane. 

Let us now apply the symmetry unitary operation
$\exp[i\frac{\pi}{4}(\sigma^{z}+1)(\tau^{z}-1)]$ to
$\sta\cs$. We, thus, obtain the state $\sta\ce$, which may be written as
\bea
\sta\ce &=&
\cos\frac{\theta(\br)}{2}
\left({1}\atop{0}\right)\otimes\left({1}\atop{-1}\right)
\nonumber\allowbreak \\
&&+
\sin\frac{\theta(\br)}{2}e^{i\phi(\br)}
\left({0}\atop{1}\right)\otimes\left({1}\atop{1}\right)\, .
\label{entText}
\eea 
In this case, one recovers the pure entanglement texture mentioned
at the beginning of Sec.~\ref{entanglement}.  As we have already
mentioned there, $\langle\sigma^{x,y}\rangle=0$ in this state whereas
the value of $\sigma^{z}$ changes just as for an ordinary $\cs$ spin
texture.  Likewise, we find that $\langle\tau^{y,z}\rangle=0$, and
thus 
\beq
\langle\tau^{x}\rangle=
-\cos\theta(\br)=-\frac{\lambda^{2}-\br^{2}}{\lambda^{2}+\br^{2}}\, .
\eeq

\subsection{General CP$^3$ Skyrmions}

Finally, it is interesting to extend this analysis to the most general CP$^{3}$
texture, subject to the condition that $\tau^x=-\sigma^z=1$ for the `point'
$|z|=\infty$. 
It is generated by the local spinor $\sta\cg$ defined by
\beq
\sta\cg=(|\lambda_{1}|^{2}+|\lambda_{2}|^{2}+2|z|^{2}+2|b|^{2})^{-1/2}
\left({{\lambda_{1}}\atop{\lambda_{2}}}\atop{{z-b}\atop{z+b}}\right)
\eeq
where $z=x+iy$, and where we denote the normalization prefactor as $\cn$
in what follows. Here $\lambda_1$, $\lambda_2$ and $b$ are complex parameters. 
The quantity $|b|^{2}+(|\lambda_1|^{2}+|\lambda_2|^{2})/2$ is the overall
length scale of the texture. The physical meaning of these parameters 
is seen more explicitly, by considering special cases, which allow one
to recover the two textures (\ref{spinText}) and (\ref{entText}) already discussed,
\beq
\sta\cs=(2\lambda_1^{2}+2|z|^{2})^{-1/2}\left({{\lambda_1}\atop{\lambda_1}}
\atop{{z}\atop{z}}\right),
\eeq
\beq
\sta\ce=(2\lambda_1^{2}+2|z|^{2})^{-1/2}
\left({{\lambda_1}\atop{-\lambda_1}}\atop{{z}\atop{z}}\right),
\eeq
and the abovementioned bimeron texture
\beq
\sta\cm=(2b^{2}+2|z|^{2})^{-1/2}\left({{0}\atop{0}}\atop{{z-b}\atop{z+b}}\right).
\label{eq:cm}
\eeq
Further insight may be obtained by considering the effect of a uniform rotation
by an angle $\gamma$ of the spin around the $z$-axis. This changes
$(\lambda_1,\lambda_2,b)$ into
$(e^{-i\gamma}\lambda_1,e^{-i\gamma}\lambda_2,b)$, after multiplying the spinor
by a global phase that brings it back to the above form. Similarly, a uniform
rotation along the $x$-axis in pseudospin space can be represented by the
matrix 
\beq
\left(\begin{array}{lcr}
e^{i\frac{\chi}{2}}\cos\frac{\chi}{2} & -i
e^{i\frac{\chi}{2}}\sin\frac{\chi}{2} & 0 \\ 
-i e^{i\frac{\chi}{2}}\sin\frac{\chi}{2} &
e^{i\frac{\chi}{2}}\cos\frac{\chi}{2} & 0 \\ 
0 & 0 & e^{i\chi}
\end{array}
\right)
\eeq
acting on the column vector $(\lambda_1,\lambda_2,b)^{T}$.

What are the entanglement properties of $\cg$? We define 
$2\lambda=\lambda_1+\lambda_2$ and $2\delta=\lambda_1-\lambda_2$, so 
that we obtain
\bea
\Xi=\frac{16}{\cn^4}
\left[
|\delta^2 z^2|+|\lambda^2 b^2|+2|\lambda\ \delta\  z\ b|\cos\phi
\right]\ ,
\eea
where $\phi=\arg(z\,\bar{b}\,\bar{\lambda}\,\delta)$. 

Thence, there are always two unentangled points, $\Xi=0$, the one by
construction at $|z|=\infty$ as well as one at $z_m=-b\lambda/\delta$, at which
the spinor is given by
$$\frac{1}{\cn}\left({1}\atop{-\frac{2b}{\lambda_1-\lambda_2}}\right)\otimes
\left(\lambda_1\atop{\lambda_2}\right).$$

To maximize the entanglement, let us first consider $|z|$  to be fixed. The
above expression shows that we should pick $\phi=0$, which means that $z$
should belong to the line passing through $z_m$ and the origin, in the
direction opposite to $z_m$. It turns out that if $z$ moves away from the
origin along this half-line, $\Xi$ first increases, reaches a maximum for
$|z|=|z_M|$ and then decreases. $|z_M|$ is given by: 
\bea
|z_{M}|=-\frac{|\lambda b|}{|\delta|}+
\sqrt{(|b|^2+|\delta|^2+|\lambda|^2)+|b|^2|\lambda|^2/|\delta|^2}\ ,
\eea
which moves to $\infty$ for $\delta=\lambda_1-\lambda_2=0$.

From this expression, one can read off the following special cases.
The {\em only} completely unentangled textures, $\Xi(z)\equiv0$, are
those with $\delta=0$ and either $b=0$ or $\lambda=0$. The former
corresponds to the Skyrmion $\cs$; the latter to the bimeron $\cm$.

Complete entanglement at  a point, $\Xi(z_M)=1$, is achieved only
in the case of $|\lambda|=|b|$, for any $\delta$. The case $b=0$
corresponds to the entanglement Skyrmion $\ce$ introduced above. The
special feature of $b=0$ is that maximal entanglement is obtained not just
at two isolated points but on the circle $|z_M|=|\lambda_1|$. In that
sense, $\ce$ is the most entangled texture.

We may, furthermore, study the  `entanglement operator'
\bea\label{eq:EOP}
\Gamma(\gamma)=\exp[i\gamma\tau^z(1+\sigma^z)]\ ,
\eea
which transforms $\cg$ into $\cg^\prime$ by replacing
\bea
\lambda_{1}^\prime=\lambda_{1}\exp[2i\gamma]\ , \qquad
\lambda_{2}^\prime=\lambda_{2}\exp[-2i\gamma]\ .
\eea
Otherwise, the structure of the relevant equations remains 
unchanged, and $\lambda$ and $\delta$ become
\bea
\lambda^\prime &=& \lambda\cos(2\gamma)+i \delta\sin(2\gamma)\nonumber\\
\delta^\prime &=& \delta\cos(2\gamma)+i \lambda\sin(2\gamma)\nonumber\ .
\eea
These equations show that $\lambda'$ moves along an elliptical orbit in the complex
plane. In the case where the modulus of $b$ lies between those of the
major and minor axes of the ellipse ($\lambda$ and $\delta$ if they
are real), there will thus be a value of $\gamma$ for which maximal
entanglement occurs.

\section{Energy scales and anisotropies in bilayers QH systems and graphene}

\subsection{Bilayer Hamiltonian}
In terms of the density operators defined in App.~\ref{app:LLLalg}, the
Hamiltonian for a bilayer quantum Hall system, taking into account both spin
and pseudospin degrees of freedom, reads
\bea\label{bilayerH}
\nn
H&=&\frac{1}{2}\sum_{\bq}V_+(\bq)\rhobar(-\bq)\rhobar(\bq)+2\sum_{\bq}V_-(\bq)
\Pbar^z(-\bq)\Pbar^z(\bq)\\
\nn
&&+\Delta_Z\Sbar^z(\bq=0)+\Delta_{T}\Pbar^x(\bq=0)+\Delta_I\Pbar^z(\bq=0),\\
\eea
where the first two terms are due to the different intra- and interlayer 
electron-electron interactions, with 
$$V_{\pm}(\bq)=\frac{\pi e^2}{\epsilon q}e^{-q^2/2}\left(1\pm e^{-qd}\right),$$
in terms of the layer separation $d$ and the dielectric constant
$\epsilon$. The last three terms in Eq. (\ref{bilayerH}) are Zeeman-type terms;
the first one is indeed the Zeeman term that acts on the $z$-component of the
spin density, whereas the second one is due to interlayer tunneling, and the
last one is a layer imbalance term which may be tuned by applying a gate
voltage. Whereas the Zeeman-type terms break the SU(4) symmetry explicitly,
the first interaction term is SU(4)-symmetric, and the second one gives rise to
an easy-plane anisotropy for the pseudospin magnetization, and thus breaks the
SU(2) pseudospin symmetry down to U(1).

In the case of GaAs heterostructures, we have the energy scales (for a
dielectric constant of $\epsilon\sim 13$ and an effective spin coupling
$g=-0.44$) 
\begin{center}
\vspace{0.2cm}
\begin{tabular}{|c||c|c|}
\hline
energy & value for arbitrary $B$ & value for $B=6$T ($\nu=1$)\\
\hline \hline
$e^2/\epsilon l_B$ & $50\sqrt{B{\rm [T]}}$K & $122$K \\ \hline
$\Delta_{Z}$ & $0.33 B{\rm [T]}$K & $2$K \\ \hline
$\Delta_{T}$ & -- & $0\, ...\, 100$K\\
\hline
\end{tabular}
\vspace{0.2cm}
\end{center}
where the last term is sample dependent. It can, for instance, be varied by
considering samples with different distance between the layers.
Indeed, the tunneling term $\Delta_T$ can vary over a rather large range and -- most
importantly -- it may become the smallest of these energy scales. One notices
that the largest energy scale is given by the interaction energy. Strictly
speaking, the characteristic intralayer interaction is given by $e^2/\epsilon
l_B$, whereas the interlayer correlations are governed by $e^2/\epsilon
\sqrt{d^2+l_B^2}$. However, we typically have $d\sim l_B=26/\sqrt{B{\rm [T]}}$ nm. 
The Zeeman term is
roughly two orders of magnitude smaller than the characteristic interaction
energy. This means that although both the Zeeman effect and interlayer
tunneling tend to fully polarize the system, with the spin aligned along the
$z$-axis and the pseudospin along the $x$-axis, this polarization only occurs
subsequent to the onset of ferromagnetic order.

Notice that the Hamiltonian (\ref{bilayerH}) also applies to the case of
graphene, a system which is, however, less anisotropic than bilayer GaAs
heterostructures. The interactions are dominated by the symmetric part,
$V_+(\bq)$, of the interaction, whereas symmetry-breaking interactions are
suppressed by a factor $a/l_B$, in terms of the carbon-carbon distance $a=0.14$
nm,\cite{GMD,alicea,abanin} which replaces $d$ as the second characteristic
length scale. As for GaAs heterostructures, the Zeeman effect in graphene plays
a minor role as compared to the leading interaction energy scale. Concerning
possible Zeeman-type terms acting on the pseudospin, which describes the two
inequivalent Dirac points, their existence remains an open and intense field of
research.\cite{fuchs,herbut,lukyanch} However, these terms are expected to be
also small with respect to the interaction energy. The energy scales for
graphene are summarized in the following tabular (for typical values
$\epsilon\sim 2.5$ and $g\sim 2$)
\begin{widetext}
\begin{center}
\vspace{0.2cm}
\begin{tabular}{|c||c|c|c|}
\hline
energy & value for arbitrary $B$ & value for $B=6$T & value for $B=25$ T\\
\hline \hline
$e^2/\epsilon l_B$ & $250\sqrt{B{\rm [T]}}$K & $620$K & $1250$K \\ \hline
$\Delta_{Z}$ & $1.2 B{\rm [T]}$K & $7$K & $30$K \\ \hline
$\Delta_{sb}<(e^2/\epsilon l_B)(a/l_B)$ & $<1 B[{\rm T}]$K & $<6$K & $<25$K\\
\hline
\end{tabular}
\vspace{0.2cm}
\end{center}
\end{widetext}
We have displayed the values both for $6$T and for $25$T because the filling
factor $\nu=1$ may be obtained for different values of the magnetic field,
due to a control of the carrier density by application of a gate voltage. The
latter choice ($25$T) is motivated by the fact that above this value, the
$\nu=1$ QH plateau is well developed.\cite{zhang}

\subsection{Physical relevance of entanglement in bilayer systems and graphene}

\begin{figure}
\centering
\includegraphics[width=7.5cm,angle=0]{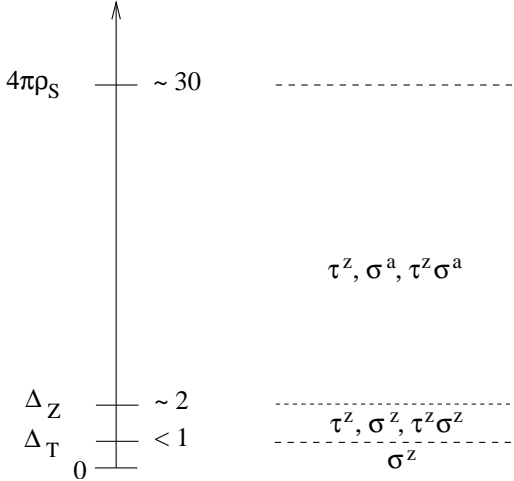}
\caption{\footnotesize{Energy scales for GaAs bilayer QH system for a magnetic 
field of $B=6$ T. The spin stiffness is given by
$\rho_S=(16\sqrt{2\pi})^{-1} e^2/(l_B\epsilon)$.
On the RHS, we indicate the symmetry generators for the Hamiltonian obtained by 
neglecting terms with energy scales below $k_B T$. 
}}
\label{fig02}
\end{figure}

The above considerations and the possibility of entangled texture states 
sensitively depend on the smallness of the tunneling gap in bilayer systems.
Experimentally, a small tunneling gap is a reasonable assumption as discussed
above, and we consider the case $\Delta_T=0$ in the following.

In the temperature regime $T<\Delta_Z$, one may then find entangled
states with an internal symmetry spanned by the Cartan algebra 
$\{\tau^z,\sigma^z,\tau^z\sigma^z\}$ of the SU(4) group. As shown above, 
the degeneracy supports entangled texture states in this regime. In the absence
of $\Delta_Z$, further entangled states arise, described by an SU(2) $\times$
U(1) symmetry group, which is larger and non-Abelian -- the operators
$\{\tau^z,\sigma^a,\tau^z\sigma^a\}$, do not all commute due to the
algebra satisfied by the Pauli matrices $\sigma^a$ (see Appendix A). This means
that one may chose any $\sigma^a$, instead of just $\sigma^z$,
to create entangled states, with the help of the entanglement operator
$$
\Gamma(\gamma^a)=\exp[i\gamma^a\tau^z(1+\sigma^a)],
$$
which generalizes that in Eq. (\ref{eq:EOP}). A full analysis of Skyrmions in
this regime is deferred to future work.

\begin{figure}
\centering
\includegraphics[width=7.5cm,angle=0]{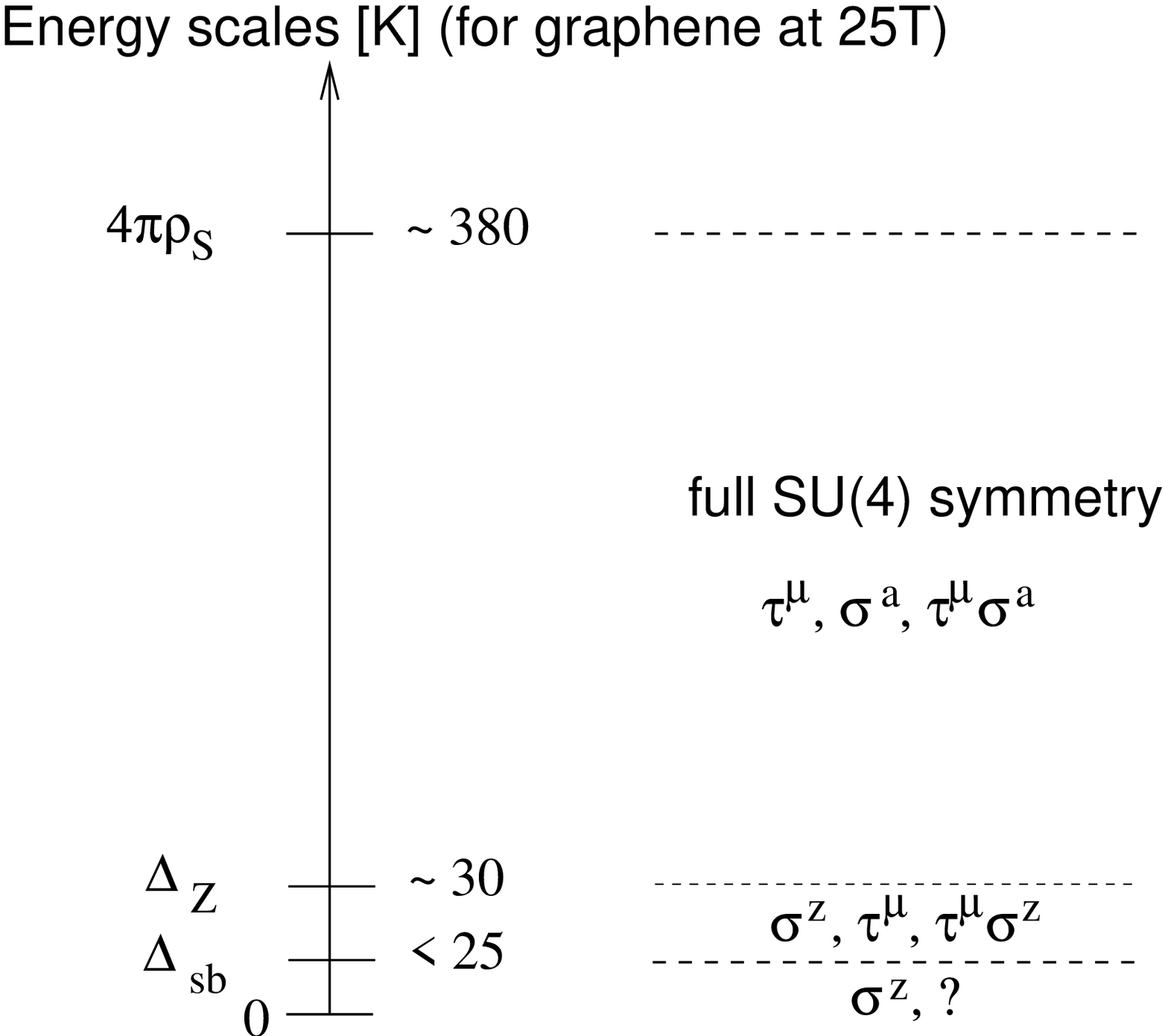}
\caption{\footnotesize{Energy scales for graphene in $B=25$ T. On the RHS, we indicate the symmetry generators for the Hamiltonian obtained by neglecting terms with energy scales below $k_B T$. 
}}
\label{fig03}
\end{figure}

The graphene case is quite similar. The energy scales are shown in Fig.
\ref{fig03}, for a characteristic field of $B=25$ T, where the $\nu=1$ 
plateau in the Hall resistance is well established.\cite{zhang}
The interaction potential has been shown to be SU(4) symmetric over a 
large temperature range, and a broken SU(2) pseudospin symmetry 
due to lattice effects is expected only below an energy scale
$\Delta_{sb}\sim (e^2/\epsilon l_B)(a/l_B)\sim B[{\rm T}]$ 
K.\cite{GMD,alicea,abanin,herbut} The nature of this symmetry breaking differs
in the relativistic LLs $|n|\geq 1$ from that in the central one $n=0$.
For $|n|\geq 1$, one expects an easy-plane anisotropy,\cite{GMD}
which supports entangled states with an internal symmetry generated by
$\{\tau^z,\sigma^z,\tau^z\sigma^z\}$, similarly to the bilayer case in the
regime $\Delta_T < T < \Delta_Z$. The $n=0$ case is more involved, and the
easy-plane anisotropy of the pseudospin magnetization\cite{GMD,abanin} is in
competition with an easy-axis anisotropy,\cite{alicea,herbut} which may lead to
a residual ${\rm Z}_2$ pseudospin symmetry. However, in all cases numerical
prefactors strongly suppress the energy scale for lattice effects 
$\Delta_{sb}\sim (e^2/\epsilon l_B)(a/l_B)$ below the typical Zeeman 
energy scale $\Delta_Z\sim 1.2B[{\rm T}]$ K. 

In an intermediate regime of $\Delta_{sb} < T < \Delta_Z$, the internal 
symmetry is given by $\{\tau^{\mu},\sigma^z,\tau^{\mu}\sigma^z\}$, which
supports entangled texture states. For temperatures well above $\Delta_Z$,
one recovers in turn the full SU(4) symmetries, and SU(4)
Skyrmions\cite{ezawa1,arovas,yang} are expected to be the relevant charged
excitations. 

\section{Dynamical properties of entangled textures}

One strong experimental motivation for the present work has been the experimental
observation of nuclear spin relaxation~\cite{Spielman05,Kumada05} in the vicinity of the 
transition between the quantum Hall pseudospin ferromagnetic phase and the compressible 
phase in a bilayer system at total filling factor $\nu=1$, where the control parameter 
is the ratio $d/l_B$. These two studies agree on the fact that nuclear relaxation is much 
faster in the compressible phase than in the quantum Hall phase. What is particularly striking 
is that a slight deviation of the total filling factor away from $\nu=1$ in the quantum Hall
phase induces a significant increase in the relaxation rate $1/T_1$. Such a strong variation 
with the filling factor is reminiscent of what has been observed in a monolayer at 
$\nu=1$,\cite{barretNMR} where the additional relaxation is attributed to the gapless spin modes
induced by the presence of Skyrmions.\cite{Cote97} In the context of a bilayer, this was surprising,
because the samples are widely believed to be fully spin-polarized in their quantum Hall phase at
$\nu=1$, and therefore the charged textures would be pseudospin bimerons, which do not couple to nuclear
spins. The dominant interpretation of these data is based on the idea of a coexistence between
both quantum Hall and compressible phases in the vicinity of the transition.\cite{Spielman05,Kumada05}
This idea has been first proposed~\cite{Stern02} to explain the strong longitudinal Coulomb drag peak
observed in the vicinity of the transition~\cite{Kellog} and is supported, on the theoretical side,
by numerical diagonalization studies~\cite{Schliemann01} which point towards a first order transition.
Furthermore, some recent experiments have confirmed the fact that there exists a large jump in the
spin polarization between the two phases.\cite{Giudici08}

Here we would like to revive the case for another scenario. It is not clear to us why the
compressible phases found either at $\nu=1$ and large $d/l_B$ or  away from $\nu=1$ for 
smaller $d/l_B$ should have the same spin excitations. This assumption is supported by
experimental results~\cite{Spielman05,Kumada05}  which show a comparable $1/T_1$ in both cases
(see for instance figures 3(b) and 3(c) in Ref.~\onlinecite{Spielman05}). But theoretically, we may
question why composite fermion physics~\cite{HLR} which is expected to emerge at large
$d/l_B$ with each layer at half filling, should remain relevant at small $d/l_B$ and away
from $\nu=1/2$ per layer. Note however that numerical\cite{Simon03,Moller07} 
and analytical\cite{Papic07} calculations indicate that, 
even for a homogeneous system, a small but non-zero layer separation admixes 
composite-fermion components to the incompressible superfluid state.
An alternative way to account for the strong relaxation away from $\nu=1$
in the gapped phase is to assume the presence of entangled textures. On purely energetic grounds,
the pure spin texture is not favorable, because of the large Zeeman energy. We may then expect
that an entangled texture would have a significantly lower energy than a spin texture, and that it
can still contribute to nuclear spin relaxation. It has in fact been shown that a lattice of entangled
skyrmions has even a lower energy than a lattice of spin-polarized bimerons if $d/l_B$ is large
enough and the Zeeman coupling is not too large.\cite{Bourassa06} Leaving aside these delicate
energetic questions, which require to optimize the CP$^{3}$ profile in the presence of  
the various anisotropies,~\cite{tsitsishvili,Bourassa06} we now briefly address the question of the collective
excitations around an entangled texture, and their impact on electronic spin-spin correlation
fucntions. Note that detailed numerical calculations of the collective modes for CP$^{3}$ skyrmion 
crystals are available.\cite{Cote07} Here, we would like simply to present a simple framework
to analyze the small fluctuations around an entangled texture, focussing on the counting of zero modes.
A striking result is that the NMR relaxation rate
induced by spin fluctuations around a given texture is changed when this
texture is modified by the application of the entanglement operator
$\exp(i\gamma\sigma^{z}\tau^{z})$, even when this operator is an exact symmetry
of the Hamiltonian, i.e. when $\Delta_{T}$ [see Eq. (\ref{bilayerH})] vanishes. 

For simplicity, we consider the SU$(4)$ symmetric case. In this
situation, the general texture that carries a unit topological charge has the
form\cite{Rajaraman}
\beq
\vec{w}(z)=z\vec{u}+\vec{v},
\eeq
where $\vec{w}(z)$ parameterizes in fact the ray in CP$^{3}$ that contains
this vector. Because both $\vec{u}=(u_1,u_2,u_3,u_4)$ and
$\vec{v}=(v_1,v_2,v_3,v_4)$ have four complex components, we have a seven
complex parameter family of degenerate textures, after factoring out the
simultaneous multiplication of $\vec{u}$ and $\vec{v}$ by the same arbitrary
complex number. We therefore expect seven zero modes for the dynamics of small
fluctuations around such texture. Note that in the absence of the texture,
$\vec{w}(z)$ is constant, so there are three zero modes that may be attributed
to the ferromagnetic vacuum sector and four to the texture: one of them
corresponds to a translation ($z\rightarrow z+a$) and the three remaining ones
to the three independent directions in the internal space of CP$^{3}$.

It is in fact illuminating to consider the case of $q$ textures. Then, the minimal
energy configurations are described as\cite{Rajaraman}
\beq
\vec{w}(z)=\sum_{j=0}^{q}z^{j}\vec{u}_{j},
\eeq
where the $N$-component $q+1$ complex vectors $\vec{u}_{j}$ are arbitrary. We have therefore
a highly deformable system where $(q+1)N-1$ degrees of freedom exhibit no restoring
force. For clarity, we are generalizing here to the CP$^{N-1}$ model with global SU($N$)
symmetry, $N$ being equal to four in the case of a bilayer. Removing the $N-1$ degrees
of freedom associated with the ferromagnetic vacuum sector leaves $qN$ of them attached to the
$q$ textures, so finally, there are $N$ static degrees of freedom per texture.
Note that the SU($N$) symmetric CP$^{N-1}$ model describes a very special system, where
the textures do not interact. In reality, various anisotropies and also the long-range
Coulomb interaction between the electric charges bound to the textures lift this massive
degeneracy within this ideal $q$-skyrmion manifold, stabilizing
most likely a crystal of skyrmions. Without anisotropies, nor Coulomb repulsion,
the $qN$ static degrees of freedom describing the small deformations around such crystal
can be organized into $N$ non-dispersive Bloch bands. In the presence of realistic
anisotropies and the Coulomb repulsion, these bands acquire a dispersive character.
Clearly, one of these $N$ branches of the collective spectrum will be a magnetic-field phonon mode,
with\cite{magnetophonon} $\omega \simeq |q|^{3/2}$, and $N-1$ other branches will be
Goldstone-type modes. We therefore recover the counting that emerged from a more microscopic
analysis.\cite{yang}

Let us now return to the quantum dynamics of the electron system around a single texture.
As in Refs. \onlinecite{Cote91,Cote97,Ezawa02,Cote07}, 
we consider the framework of the time-dependent Hartree-Fock (TDHF)
approximation, which can be formulated in several different ways. A rather
appealing one consists of viewing the set of Slater determinants as a classical
phase space.\cite{Rowe80} The Hartree-Fock Hamiltonian is then 
used as a classical Hamiltonian that generates the same dynamics
as the TDHF equations of motion. In our problem of a filled LL,
we also assume (see the beginning of Sec. II) that the occupied
single-particle states correspond to localized four-component spinors
$w_{j}(\br)$ that are {\em not} normalized here. Indeed, dropping the
normalization constraint simplifies the equation of motion at the expense of
introducing a form of a generalized local gauge symmetry $w_{j}(\br)\rightarrow
h(\br)w_{j}(\br)$  which has to be factored out in the overall mode
counting. The Hartree-Fock Hamiltonian may be taken [assuming SU$(4)$ symmetry]
as 
\beq
H(\{w\})=\frac{1}{4}\int d^2r\left(\frac{(\nabla w,\nabla w)}{(w,w)}-
\frac{(\nabla w,w)(w,\nabla w)}{(w,w)^{2}}\right),
\eeq 
where the inner product is defined by
$(w,w')=\sum_{j=1}^{4}\bar{w}_{j}(\br)w'_{j}(\br)$. This is the only local Hamiltonian 
that is quadratic in gradients, invariant under the
global SU$(4)$ transformations in spinor space and also under local gauge
transformations $w_{j}(\br)\rightarrow h(\br)w_{j}(\br)$.\cite{Rajaraman}
As usual, it is now convenient to introduce the complex position coordinate
$z=x+iy$, such that
\bea
\nn
H(\{w\})&=&\frac{1}{2}\int d^2r\left(\frac{(\bar{\partial}w,\partial
  w)+(\partial w,\bar{\partial}w)} {(w,w)}\right.\\
\nn
&&-\left.\frac{(\bar{\partial}w,w)(w,\partial w)+(\partial
  w,w)(w,\bar{\partial}w)}{(w,w)^{2}}\right). 
\eea
The notations mean in particular that $(\partial
w,w')=\sum_{j}(\partial_{z}\bar{w}_{j})w'_{j}$ and $(\bar{\partial}
w,w')=\sum_{j}(\partial_{\bar{z}}\bar{w}_{j})w'_{j}$. The classical Hamilton
equations of motions are 
\bea \label{eqmotion}
\nn
i\frac{\partial}{\partial t}w_{j}(\br,t) & = & \frac{\delta H}{\delta
  w_{j}(\br)}\, , \\ 
i\frac{\partial}{\partial t}\bar{w}_{j}(\br,t) & = & -\frac{\delta H}{\delta
  \bar{w}_{j}(\br)}\, . 
\eea
Because of the local gauge invariance, the norm $(w(\br),w(\br))$ is conserved
for any $\br$. The price to pay for this is that the equations of motion are rather
complicated. Using the gauge symmetry, we may modify Eq.~(\ref{eqmotion}) by a
term proportional to $w_{j}(\br)$ that simplifies it considerably, giving up
the conservation of $(w(\br),w(\br))$, while preserving the ray of $w(\br)$ in
CP$^{3}$ and therefore the physical content of the solutions. This simplified
equation of motion reads 
\beq
i\frac{\partial}{\partial
  t}w=-\partial^{2}_{z,\bar{z}}w+\frac{(w,\partial_{\bar{z}}w)}{(w,w)}\partial_{z}w +\frac{(w,\partial_{z}w)}{(w,w)}\partial_{\bar{z}}w+\lambda(\br,t)w\, , 
\eeq
where $\lambda(\br,t)$ is an arbitrary complex function of $\br$ and $t$, and
$\partial^{2}_{z,\bar{z}}\equiv \partial_z\partial_{\bar{z}}$.
Note that the gauge transformation $w'(\br,t)=h(\br,t)w(\br,t)$ sends a
solution $w$ into another solution $w'$ provided $\lambda$ is changed into
$\lambda'$ given by
\bea
\nn
\lambda' &=& \lambda+h^{-1}\left[i\frac{\partial h}{\partial
	t}+\partial^{2}_{z,\bar{z}}h
  -\frac{(w,\partial_{\bar{z}}w)}{(w,w)}\partial_{z}h\right. \\
\nn
&&-\left.\frac{(w,\partial_{z}w)}{(w,w)}\partial_{\bar{z}}h\right]
-2\frac{\partial_{\bar{z}}h\partial_{z}h}{h^{2}}
\eea
It is immediate to check that the general static texture $w(\br)=zu+v$
is indeed a solution with $\lambda=0$.

Linearizing the equations of motion around these static solutions gives
\beq
i\frac{\partial}{\partial t}\epsilon=-\partial^{2}_{z,\bar{z}}\epsilon+
\frac{(w,\partial_{\bar{z}}w)}{(w,w)}\partial_{z}\epsilon
+\frac{(w,\partial_{z}w)}{(w,w)}\partial_{\bar{z}}\epsilon+\lambda(\br,t)w\, ,
\label{linear_eqmotion}
\eeq  
where $\epsilon(\br,t)$ denotes an infinitesimal variation and we have used the
fact that $\partial_{\bar{z}}w=0$ for the static texture. Note that $\lambda$
is an infinitesimal complex function of the same order as $\epsilon$.
A first order infinitesimal gauge transformation with $h(\br,t)=1+\zeta(\br,t)$
acts as
\bea 
\nn
\epsilon' & = & \epsilon+\zeta w\ , \\
\nn
\lambda' & = & \lambda+i\frac{\partial \zeta}{\partial t}+\partial^{2}_{z,\bar{z}}\zeta
-\frac{(w,\partial_{z}w)}{(w,w)}\partial_{\bar{z}}\zeta\, .
\eea
The eigenmodes have the form $\epsilon(\br,t)=\epsilon(\br)e^{-i\omega
  t}$. Because the classical Hamiltonian is bounded below, $\omega$ can take
only non-negative values. The zero modes are obtained from the condition
$\partial_{\bar{z}}\epsilon=0$, and the constraint of a fixed topological
charge imposes that $\epsilon$ should be a polynomial of degree 1 in $z$. This
yields an eight-dimensional space of zero modes in which the pure gauge
transformations $\epsilon=\zeta_{0}w$ with $\zeta_{0}$ constant in space and
time, have to be removed. We therefore recover the seven dimensional complex
space of zero modes discussed at the beginning of this section.

Note that we have a conserved Hermitian form $\Omega(\eta,\xi)$ for any pair
$\eta, \xi$ of solutions of the linearized equations of
motion~(\ref{linear_eqmotion}). This form $\Omega$ is defined by
\beq
\Omega(\eta,\xi)=\int d\br \frac{(\eta_{\perp},\xi_{\perp})}{(w,w)}=
\int d\br
\left(\frac{(\eta,\xi)}{(w,w)}-\frac{(\eta,w)(w,\xi)}{(w,w)^{2}}\right) 
\eeq
and is invariant under any local gauge transformation that affects
simultaneously $w, \eta$, and $\xi$. In particular, for eigenmodes
$\eta(\br,t)=\eta(\br)e^{-i\omega_{\eta} t}$ and
$\xi(\br,t)=\xi(\br)e^{-i\omega_{\xi} t}$, we have $\Omega(\eta,\xi)=0$ as soon
as $\omega_{\eta}\neq \omega_{\xi}$. This shows that a complete eigenmode basis
$\epsilon^{\alpha,\omega}(\br)$ ($\alpha$ being an internal polarization label)
can be chosen to be orthogonal for the Hermitian form $\Omega$.

We will not try to find the full continuous spectrum of the scattering problem
defined by Eq.~(\ref{linear_eqmotion}). Instead, we shall now consider the
effect of the entanglement operator $\exp(i\gamma\sigma^{z}\tau^{z})$ on the
local spin correlation function $C(\br,\omega)=\int dt\langle
S^{+}(\br,t)S^{-}(\br,t=0)\rangle e^{i\omega t}$ that controls the local
nuclear spin relaxation rate at the position $\br$. Here,
$S^{\pm}(\br,t)=S^{x}(\br,t)\pm i S^{y}(\br,t)$, in terms of the local spin
densities $S^{x/y}(\br,t)$. In the presence of a texture, translation
invariance is lost, so this correlation function will explicitly depend on
$\br$. The TDHF approximation does not give us directly wave-functions for
collective modes, so we use it to express the corresponding response function 
$R(\br,\omega)=\frac{i}{\hbar}\int dt\langle
[S^{+}_{\br}(t),S^{-}_{\br}(0)]\rangle e^{i\omega t}$ in terms of a complete
basis of eigenmodes $\epsilon^{\alpha,\omega}$. $R(\br,\omega)$ is related to
the correlation function by the usual fluctuation-dissipation formula,
\beq
C(\br,\omega)=\frac{2\hbar}{1-e^{-\beta \hbar \omega}}\Im R(\br, \omega)\, .
\eeq
Introducing the $4\times4$ matrices $M^{\pm}=\sigma_{\mathrm{spin}}^{\pm}
\otimes \bone_{\mathrm{iso}}$, the response function is then expressed as
\begin{widetext}
\beq
\Im R(\br, \omega)  =  \pi
\sum_{\alpha}\frac{(w(\br),M^{+}\epsilon_{\perp}^{\alpha,\omega}(\br)) 
(\epsilon_{\perp}^{\alpha,\omega}(\br),M^{-}w(\br))}
	{(w(\br),w(\br))^{2}}\theta({\omega}) 
-\pi  \sum_{\alpha}\frac{(w(\br),M^{-}\epsilon_{\perp}^{\alpha,-\omega}(\br))
(\epsilon_{\perp}^{\alpha,-\omega}(\br),M^{+}w(\br))}
{(w(\br),w(\br))^{2}}\theta({-\omega})\,  
. 
\label{sum_modes}
\eeq
\end{widetext}

Notice that we have treated, within the calculation of the low-energy modes,
the Skyrmion as a classical object, which is a valid approximation for large
Skyrmions, $\lambda\gg l_B$. For smaller Skyrmions, the translational modes may
still be treated on the classical level, whereas the rotational modes need to
be quantized.\cite{GirvinHouches} 

Now, let us start from the pure spin texture discussed in Sec. IV and let us
apply the entanglement operator $\exp(i\gamma\sigma^{z}\tau^{z})$ to it. This
unitary transformation in the full quantum Hilbert space induces a canonical
transformation on the classical phase-space of Slater determinants. It is,
therefore,
equivalent to compute the correlation function $\langle S^{+}S^{-}\rangle$
around the transformed texture and the correlation function $\langle
S^{+}(\gamma)S^{-}(\gamma)\rangle$ around the original one, where
$\bs(\gamma)=\exp(-i\gamma\sigma^{z}\tau^{z})\bs\exp(i\gamma\sigma^{z}\tau^{z})$.
Explicitly, we obtain 
\beq
S^{\pm}(\gamma)=\cos(2\gamma)S^{\pm}\otimes\bone\mp
i\sin(2\gamma)S^{\pm}\otimes\tau^{z}. 
\eeq
For the fluctuations around the pure spin texture, a little inspection
shows that the $\tau^{x}$ component of pseudospin is conserved. Therefore,
crossed terms involving only one $\tau^{z}$ operator do not contribute to the
sum over eigenmodes~(\ref{sum_modes}). We finally get

\bea
\langle S^{+}S^{-}\rangle_{\gamma} &=& \cos^{2}(2\gamma)\langle
S^{+}S^{-}\rangle_{\gamma=0}\nonumber \\ 
&&+ \sin^{2}(2\gamma)\langle (S^{+}\tau^{z})(S^{-}\tau^{z})\rangle_{\gamma=0}~~~.
\label{eq44}
\eea

Note that the two correlation functions that appear on the RHS of the above equation
will in general differ, because we are considering \textit{local} spin and pseudospin operators.
So the two $\tau^{z}$ operators taken at two different times are not expected to commute,
in spite of the fact that the $z$-component of the \textit{total} pseudospin is conserved.
This expression is quite interesting because it exhibits a competition between
two effects. When $\gamma$ increases slightly away from zero, the first term
decreases, in agreement with physical intuition -- the local spin is reduced by
entanglement, and so is its contribution to NMR relaxation. Such scenario would be
consistent with the experimental observation that $1/T_1$ increases faster for a monolayer
than for a bilayer as the filling factor is moved away from the $\nu=1$ value (see for instance
Fig. 4 in~\onlinecite{Kumada05}). But this is not the
only effect induced by entanglement -- the second term also builds up, showing
that new relaxation channels occur, that correspond to flipping simultaneously
both spin and pseudospin degrees of freedom. 

A physical consequence of this variation of $1/T_1$ with the entanglement parameter $\gamma$
could be the existence of large fluctuations in the nuclear relaxation rate from one sample to
another, or for the same sample after thermal cycling between the low temperature quantum Hall state
and the normal electron fluid. Indeed, we may speculate that, in the vicinity of the bilayer phase transition, 
the energy landscape in the space of possible textures is characterized by a rather flat minimum around 
the bimeron texture, and that it may then be difficult for the system to relax to its absolute ground-state.
If this picture were to be confirmed, the above discussion shows that $1/T_1$ would depend on the
actual degree of entanglement reached by the system.

\section{Summary and Outlook}

In view of the intense current interest in the physics of entanglement -- in part motivated by the experimental advances in cold atoms, the search for exotic electronic phases and attempts to build a quantum computer -- quantum Hall systems present and attractive subject of study, not least given the presence of several phases exhibiting macroscopic quantum coherence.

We have investigated the possible entanglement between two internal SU(2) degrees of freedom of
electrons restricted to a single LL. Whereas the physical spin constitutes one SU(2) copy,
the second one describes a pseudospin degree of freedom such as the layer index in bilayer QH
systems or the valley degeneracy in graphene in a strong magnetic field.
The Schmidt decomposition allows one to treat both on an equal footing, while embedding them in a larger SU(4) space.
This has enabled us to study the effects of realistic anisotropies in a transparent manner. In particular, by defining an entanglement operator which generates entire families of degenerate Skyrmions, we are able to construct wavefunctions for entanglement Skyrmions which are easy to visualize.

It has also become apparent that the NMR response of such multicomponent systems is affected by the degree of entanglement. Indeed, it remains to be seen whether this may provide an alternative to the puddle model of coexistence of a spin-polarized incompressible and a partially polarized compressible metallic phase,\cite{Stern02} which is mainly invoked in the explanation of the increased NMR relaxation rate when driving the total filling factor away from
$\nu_T=1$.\cite{Spielman05,Kumada05} Unfortunately, in graphene, where anisotropies in the valley sector are quite weak, NMR measurements turn out to be difficult because the majority ($\sim99\%$) of the carbon atoms are $^{12}$C isotopes, which are NMR inactive due to a zero nuclear spin.

This work leaves many open questions. The first is to identify precisely the
regimes in which entangled textures can be stabilized as the absolute energy
minima for charged excitations. This has been demonstrated to occur in the
presence of charge imbalance between the two layers,\cite{tsitsishvili} or even
for the balanced system, for a layer separation lower than, but close to the critical one,
provided the Zeeman energy is not too large.\cite{Bourassa06} An important issue
here would be to understand the effect of random impurities that may disorder such
lattices of entangled skyrmions and thus modify the energy balance between various
configurations. But even when the spin polarized bimeron is the most favorable texture, 
it would be very interesting to understand better the energy landscape in its vicinity,
because available numerical results suggest it may be rather flat,\cite{Bourassa06} which
raises the possibility of a slow dynamics and a lack of equilibration.
Finally, there exist certainly many other physical signatures of spin-pseudospin entanglement
besides the local reduction in the magnitude of the average spin and pseudospin vectors.
These issues are left for future studies.

\section*{Acknowledgments}

BD and MOG thank the Rudolf Peierls Centre for Theoretical Physics, Oxford
University, where much of this work was undertaken,
for hospitality. PL acknowledges hospitality of the Departamento de
Fisica (UFPE, Recife, Brazil) and the PVE/CAPES program.
MOG and PL acknowledge financial support by the
Agence Nationale de la Recherche under Grant no. ANR-06-NANO-019-03.

\appendix
\section{Lowest-LL algebra for multicomponent systems}
\label{app:LLLalg}

The local spin and pseudospin densities may be expressed in terms 
of Pauli matrices as
\beq\label{Sdens}
\Sbar^a(\br)=\frac{1}{2}f(\br)\otimes(\sigma^a\otimes\bone)
\eeq
and
\beq\label{Pdens}
\Pbar^{\mu}(\br)=\frac{1}{2}f(\br)\otimes(\bone\otimes\tau^{\mu}).
\eeq
Similarly, the total (charge) density may be written as
\beq\label{Tdens}
\rhobar(\br)=f(\br)\otimes(\bone\otimes\bone).
\eeq
Here, $f(\br)$ is the U(1) one-particle density projected to the lowest LL,
the Fourier components of which satisfy the magnetic-translation 
algebra\cite{GMP,moon}
\beq\label{comm}
[f(\bq),f(\bq')]=2i\sin\left(\frac{\bq\wedge\bq'}{2}\right)f(\bq+\bq')
\eeq
and
\beq\label{anticomm}
\{f(\bq),f(\bq')\}=2\cos\left(\frac{\bq\wedge\bq'}{2}\right)f(\bq+\bq').
\eeq
In Eqs. (\ref{Sdens})-(\ref{Tdens}), the bar indicates that the density
operators are restricted to the lowest LL. Because of Eqs. (\ref{comm}) and
(\ref{anticomm}),
the Fourier-tranformed density operators satisfy the commutation relations
\beq\label{commD}
[\rhobar(\bq),\rhobar(\bq')]=2i\sin\left(\frac{\bq\wedge\bq'}{2}\right)
\rhobar(\bq+\bq'),
\eeq

\beq\label{commSD}
[\Sbar^{a}(\bq),\rhobar(\bq')]=2i\sin\left(\frac{\bq\wedge\bq'}{2}\right)
\Sbar^{a}(\bq+\bq'),
\eeq

\beq\label{commPD}
[\Pbar^{\mu}(\bq),\rhobar(\bq')]=2i\sin\left(\frac{\bq\wedge\bq'}{2}\right)
\Pbar^{\mu}(\bq+\bq')
\eeq

\bea\label{commSS}
[\Sbar^{a}(\bq),\Sbar^{b}(\bq')]&=&\frac{i}{2}\delta^{ab}\sin\left(
\frac{\bq\wedge\bq'}{2}\right)\rhobar(\bq+\bq')\\
\nn
&&+i\epsilon^{abc}
\cos\left(\frac{\bq\wedge\bq'}{2}\right)\Sbar^{c}(\bq+\bq'),
\eea
and
\bea\label{commPP}
[\Pbar^{\mu}(\bq),\Pbar^{\nu}(\bq')]&=&\frac{i}{2}\delta^{\mu\nu}\sin\left(
\frac{\bq\wedge\bq'}{2}\right)\rhobar(\bq+\bq')\\
\nn
&&+i\epsilon^{\mu\nu\sigma}
\cos\left(\frac{\bq\wedge\bq'}{2}\right)\Pbar^{\sigma}(\bq+\bq').
\eea
These commutation relations are directly obtained from the relation
\bea\label{relation}
\nn 
[f(\bq)\otimes A,f(\bq')\otimes B]&=&
\frac{1}{2}\left([f(\bq),f(\bq')]\otimes\{A,B\}\right.\\
\nn
&&\left.+\{f(\bq),f(\bq')\}\otimes[A,B]\right),\\
\eea
for any pair of operators $A,B$. 
Furthermore, one obtains the mixed spin-pseudospin commutator,
\beq\label{commSP}
[\Sbar^{a}(\bq),\Pbar^{\mu}(\bq')]=\frac{i}{2}\sin
\left(\frac{\bq\wedge\bq'}{2}\right)
f(\bq+\bq')\otimes\left(\sigma^{a}\otimes \tau^{\mu}\right).
\eeq
Naively, one might have expected that the spin and pseudospin densities are
decoupled because of $[\sigma^{a}\otimes \bone,\bone\otimes \tau^{\mu}]=0$. 
However, the last commutation relation indicates that the spin and 
pseudospin dynamics are indeed entangled when taking into account local
variations of the texture. This is due to the projection into the 
lowest LL, which yields the noncommutativity of the orbital degrees of 
freedom [Eq. (\ref{comm})]. 
It may also be seen directly from Eq. (\ref{relation}): if we had 
$[f(\bq),f(\bq')]=0$, the anti-commutator term $\sim\{A,B\}$ would vanish, and
one would obtain $[\Sbar^{a}(\bq),\Pbar^{\mu}(\bq')]=0$.
The same argument also yields the spin-charge and the pseudospin-charge 
entanglement, revealed by Eqs. (\ref{commSD}) and (\ref{commPD}).

Eq. (\ref{commSP}) indicates that we 
generate, via the commutators, the generators $\sigma^{a}\otimes \tau^{\mu}$ which,
together with $\sigma^{a}\otimes{\bone}$ and $\bone\otimes \tau^{\mu}$, give rise to 
the larger internal symmetry group SU(4), in which the SU(2)$\times$SU(2) 
group thus 
needs to be embedded. The SU($N$) extension of the magnetic translation
group, $\mathcal{T}_M\times\rm{SU({\it N})}$, with $N\geq4$, was
indeed the starting point of several other theoretical works on bilayer 
quantum Hall systems.\cite{ezawa1,arovas,ezawa2,tsitsishvili} 
We have thus shown that a SU(4) description of the texture states is
necessary even if the ground state has the reduced 
internal symmetry SU(2)$\times$SU(2).


\end{document}